\documentclass{ws-mpla}

\newcommand{\be}{\begin{equation}}  
\newcommand{\ee}{\end{equation}}

\begin{document}

\markboth{Gero von Gersdorff}
{The MSSM on the Interval}

\catchline{}{}{}{}{}

\title{THE MSSM ON THE INTERVAL}

\author{\footnotesize GERO VON GERSDORFF
}

\address{Department of Physics and Astronomy, Johns Hopkins University, 3400 North Charles Street\\
Baltimore, MD 21218, US
\\
gero@pha.jhu.edu}



\maketitle

\pub{Received (Day Month Year)}{Revised (Day Month Year)}

\begin{abstract}

We review electroweak symmetry breaking in supersymmetric models with
a compact fifth dimension, the interval. We show how boundary
conditions for hypermultiplets can be obtained dynamically by brane
mass terms, and present formulae for the spectrum in the presence of
general bulk mass matrices. After giving a brief overview on the
literature of models, we describe in detail a recently proposed model
that at energies below the compactification scale reduces to the MSSM
with a very peculiar superpartner spectrum.

\end{abstract}

\ccode{PACS Nos.: 12.60.-i, 12.60.Jv, 12.90.+b}

\section{Introduction}

Supersymmetric models in five dimensions with a low comapctification
scale (i.e. TeV or multi TeV) have received considerable interest in
recent years.\cite{Antoniadis:1990ew}
$^-$\cite{Diego:2006py}
The four-dimensional supersymmetric Standard Model (MSSM) naturally
incorporates electroweak symmetry breaking (EWSB), as loops of the
top/stop sector drive the Higgs mass squared to negative values, thus
enforcing the Higgs to acquire a nontrivial vacuum expectation value
(VEV). The size of the electroweak scale is controlled by the scale of
supersymmetry (SUSY) breaking, in particular the mass of the stop
itself.  The quadratic sensitivity of the electroweak scale to the
ultraviolet physics is cut off above the stop mass, and as long as the
latter does not exceed several TeV, EWSB can occur in a fairly natural
manner. In fact, understanding the mechanism that triggers SUSY
breaking is one of the main issues in supersymmetric theories, and it
should determine the phenomenology of supersymmetric particles at
future high-energy colliders as the LHC.

The standard way of beraking SUSY in models with Extra Dimensions is
the Scherk-Schwarz (SS) mechanism.\cite{Scherk:1978ta,SS2} This
breaking is unique to Extra Dimensions, as it preserves different
higher dimensional supersymetries at different points in the Extra
Dimension(s). For instance, in five dimensional (5d) models
compactified on an interval of length $\pi R$, $N=2$ SUSY is preserved
in the bulk, while at the two boundaries two different $N=1$
supercharges survive, resulting in fully broken SUSY in the 4d
effective theory.  SS-breaking exhibits very little UV sensitivity,
due to the fact that it is a nonlocal mechanism, and hence soft mass
terms corresponding to local counterterms in the 5d bulk or 4d brane
actions are not allowed. As a result, all radiative effects related to
SUSY breaking are cut off at the compactification scale. The only UV
sensitivity thus results from the supersymmetric renormalization of
the bulk and brane operators, i.e.~a supersymmetric running of the
gauge and Yukawa couplings. While the gauge coupling is always
linearly divergent, the UV-sensitivity of the Yukawa terms depend on
the nature of the chiral superfields involved. SUSY forces the Yukawa
interactions to be localized on the boundary, nevertheless they can
involve bulk fields. Simple dimensional analysis shows that the more
bulk fields enter a Yukawa interaction the more sensitive the theory
is to unknown UV physics, as wave function renornmalization of bulk
fields are linearly divergent.  Yukawa interactions with only bulk
fields also increase the number of free parameters of that model, as
one can write different couplings at the two boundaries.  While from
these considerations it seems preferable to localize all matter and
Higgs multiplets at the boundaries, it turns out that EWSB does not
take place in such a setup.\cite{Barbieri:2002sw} One is lead to
consider some matter or Higgs fields to live in the bulk. In this
case, particular care is needed in choosing the boundary conditions
for these fields in order to avoid generation of quadratically
divergent Fayet Iliopoulos (FI) terms at the
branes.\cite{Ghilencea:2001bw}

In this paper we will review several supersymmetric models of EWSB
with a fifth dimension compactified on a (flat) interval of length $\pi
R$.  Such models can be classified according to where Higgs and matter
sectors live. Fields that live in the bulk must transform as
hypermultiplets of the extended bulk supersymmetry. In
Sec.~\ref{interval} we review the description in terms of $N=1$
multiplets, relate consistent boundary conditions dynamically to
appropriate brane actions, and compute mass spectra in the presence of
general bulk mass matrices.  Successful EWSB forbids both matter and
Higgs sectors to be completely localized. In Sec.~\ref{models} we
rewiew some of the models that achieve EWSB by delocalizing either the
Higgs fields or some of the matter fields.  In Sec.~\ref{DGQ} we then
focus on a recently proposed model of a quaislocalized Higgs sector,
and describe in some detail its experimental signatures, in particular its
distinctive pattern of squark and sletpon masses.

\section{Supersymmetry on the Interval}
\label{interval}

Let us write the general Lagrangian of a bulk hypermultiplet in terms
of $N=1$ chiral superfields. The 5d vector multiplet splits into a 4d
one $V$ and a chiral adjoint multiplet $\Sigma$, while the
hypermultiplet is made up out of two chiral multplets $\mathcal H$ and
$\mathcal H^c$. The Lagrangian can then be expressed as~\cite{Marti:2001iw,Barbieri:2002ic,Marti}
\begin{multline}
\mathcal L^{\rm hyper} = \int d^4\theta\ \frac{\mathcal T+\bar
{\mathcal T}}{2} \left\{\bar{\mathcal H} \, \exp(T_a V^a) \, \mathcal
H+ \mathcal H^{c}\, \exp(-T_aV^a)\, \bar{\mathcal H}^c\right\}\\ -\int
d^2\theta \left\{\mathcal H^c (\overleftrightarrow{\partial}_y-\mathcal M \mathcal T+T_a\Sigma^a)
\mathcal H+h.c.\right\}\,,
\label{bulklag}
\end{multline}
where the $\overleftrightarrow{\partial}=\frac{1}{2}(\overleftarrow{\partial}-\
\overrightarrow{\partial})$. The mass matrix $\mathcal M$ is
hermitian and acts on the internal flavor indices of the
hypermultiplet.  The radion field $\mathcal T$ will be taken
nondynamical,
\be
\mathcal T=R+2\, \omega\,\theta^2 \,.
\ee
Its scalar component parametrizes the size of the extra dimension and
a non-zero $\omega$ implements the SS
breaking.\cite{Marti:2001iw,Kaplan:2001cg,vonGersdorff:2001ak}

Let us start with a single hypermultiplet, in which case the hermitian
matrix $\mathcal M$ is a real number that we call $M'$. The boundary
conditions can be obtained dynamically from the action principle by
adding a suitable supersymmetric brane mass term at
$y_f=0,\pi$
\be
\mathcal L_f^{\rm hyper} = -\frac{1}{2}
\int d^2 \theta\ r_f
\mathcal  H^c  \mathcal H+h.c.\,,
\label{boundcov2}
\ee
with $r_f^2=1$.\cite{Diego:2006py} The boundary conditions following from varying the action can be given in superfield form:
\be
(1- r_f)\mathcal H=0\,,\qquad  (1+ r_f)\mathcal H^c=0\,.
\label{BCs2}
\ee
As $r_f=\pm 1$, only one of the two equations in (\ref{BCs2}) is
non-trivial, giving Dirichlet boundary conditions to one chiral
multiplet, while the other superfield remains
unconstrained.\footnote{The case of gauge multiplets can be treated
analogously, see Refs.~\refcite{Belyaev:2005rs,Belyaev:2006jg}.}  In
the orbifold picture the quantities $r_f$ are known as the parities of
the field $\mathcal H$ (w.r.t.~the fixed point at $y=y_f$). The scalar
boundary conditions follow once the auxiliary fields are integrated
out:
\be
(1-r_f)H=0,\qquad (1+r_f)(\partial_y- M')H=0\,,
\label{BC3}
\ee
\be
(1+r_f)H^c=0,\qquad (1-r_f)(\partial_y+M')H^c=0\,.
\label{BC4}
\ee
For $r_0=-r_\pi=\pm 1$ one finds for the scalar mass eigenvalues
\be
\left(1+\frac{M'^2}{\Omega^2}\right) \sin^2(\Omega \pi R)
=\sin^2(\pi\omega)\,,
\label{bos++}
\ee
where the we have defined $\Omega^2=m^2-M'^2$. In the case
$r_0=r_\pi=\pm 1$ one finds
\be
\left(\cos(\pi\Omega R)\pm\frac{M'}{\Omega}\sin(\pi\Omega R)\right)^2
=\sin^2(\pi\omega)\,.
\label{bos+-}
\ee
The fermionic spectrum is obtained by setting the SS-parameter $\omega=0$.

A single hypermultiplet with these boundary conditions is known to
produce quadratically divergent FI terms for the hyperchage, localized
at the boundaries.\cite{Ghilencea:2001bw} Although they can be absorbed
by a shift in the field $\Sigma_Y$, they reappear as mass terms for
charged fields due to the Yukawa coupling in Eq.~(\ref{bulklag}). One
concludes that the mass term $M'$ is effectively renormalized and in
particular quadratically divergent. This can be avoided if two
hypermultiplets of the same kind but with ``orthogonal'' boundary
conditions are considered. To this end, one considers the two
hypermultiplets to form a doublet under a formal $SU(2)_H$ global
symmetry.\footnote{We call it a ``formal'' symmetry as it might be
explicitely broken by bulk and/or brane mass terms as well as Yukawa
interactions.}  The brane Lagrangian
\be
\mathcal L_f^{\rm hyper} = \frac{1}{2}
\int d^2 \theta\ 
\mathcal H^c\, \vec r_f\cdot\vec \sigma\, H+h.c.\,,
\label{boundcov}
\ee
produces the superfield boundary conditions
\be
(1+\vec r_f\cdot\vec \sigma)\mathcal H=0,
\qquad \mathcal H^c(1-\vec r_f\cdot\vec \sigma)=0\,,
\label{BCs}
\ee
and quadratically divergent FI terms are shown to be absent.  The two
fields in Eq.~(\ref{BCs}) that do not vanish at a given boundary carry
opposite hypercharge and can have the MSSM superpotential
couplings of the Higgs fields to boundary matter.  The misalignment of
the two vectors $\vec r_0$ and $\vec r_\pi$ can be translated into a
SS-parameter for the $SU(2)_H$ symmetry, given by $\cos(2\pi\tilde\omega)=\vec
r_o\cdot\vec r_\pi$.  The mass matrix is conveniently parametrized as
\be
\mathcal M=M'+M\, \vec p\cdot\vec \sigma\,,
\ee
where $\vec p$ is a unit vector and $M'\in \mathbb R$, $M\in \mathbb
R_+$. Again by integrating out the auxiliary fields, one finds the
bosonic boundary conditions
\be
(1+\vec r_f\cdot\vec \sigma)H=0,
\qquad 
(1-\vec r_f\cdot\vec \sigma)(\partial_y-M'+ c_f M)H=0\,,
\label{BCscalar1}
\ee
\be
H^c(1-\vec r_f\cdot\vec \sigma)=0,
\qquad 
H_c(\overleftarrow\partial_y+M'+ c_f M)(1+\vec r_f\cdot\vec \sigma)=0\,,
\label{BCscalar2}
\ee
with $c_f=\vec r_f\cdot\vec p$.  As mentioned before, the mass term
$M'$ is equivalent to a boundary $D$-term. Although no quadratic
renormalizaton of this operator occurs, there is a linear divergent
contribution $\sim M'\Lambda$. We will thus mostly be interested in
the case of $M'=0$, in which case the mass eigenvalues are given by
the zeroes of the two equations~\cite{Diego:2005mu}
\be
\left(\cos(\Omega \pi R)-\frac{c_0 M}{\Omega}\sin(\Omega \pi R)\right)
\left(\cos(\Omega \pi R)+\frac{c_\pi M}{\Omega}\sin(\Omega \pi R)\right)
=\cos^2(\omega\pm\tilde\omega)\pi\,,
\label{masses}
\ee
where now $\Omega^2=m^2-M^2$.

The generic Neumann boundary condition
\be 
(\partial_y+\Omega_f)\Phi|_{y=y_f}=0
\ee
leads to wave functions exponentially decaying in the bulk whenever
\be
\Omega_0 R \pi \gg 1\,,\qquad \Omega_\pi R \pi\ll-1\,.
\ee
These states are known as quasilocalized states. Their mass
eigenvalues follow from the spectra Eq.~(\ref{bos++}), (\ref{bos+-})
and (\ref{masses}) with $\Omega=i\Omega_f$, i.e.
\be
m_f^2= -\Omega_f^2+M^2 + \mathcal O (\epsilon)\,,\qquad \epsilon =e^{-\Omega_f \pi R}\,.
\label{locmass}
\ee
The parameter $\epsilon$ serves as an order parameter describing the
degree of quasi-localization of the corresponding state. The limit
$\epsilon\to 0$ describes a fully localized chiral multiplet while at
the same time the heavy KK modes (the ones with real $\Omega$)
decouple. Even for moderately large values of the bulk mass,
e.g. $M\sim R^{-1}$, the localization is quite efficient and
corrections to the masses Eq.~(\ref{locmass}) are strongly
suppressed. This suggests some kind of systematic expansion in the
parameter $\epsilon$.

\section{Previous Models}
\label{models}

Models of EWSB in supersymmetric theories with TeV-size Extra
Dimensions and SS-breaking can be classified acoording to where the
matter and Higgs sectors live. Any bulk multiplet will feel SUSY
breaking at the tree level, while brane fields can obtain soft masses
only through their coupling to bulk fields (i.e~the gauge sector and
possibly other matter fields) via loop effects.  A somehow hybrid
status is assumed by the quasi-localized states discussed in
Sec.~\ref{interval}. As can be seen from Eq.~(\ref{locmass}), the
leading contribution (which becomes exact in the strictly localized
limit) is independent of the SS parameter $\omega$, and all SUSY
breaking effects are controlled by the small parameter $\epsilon$.

In the limit of unbroken SUSY, EWSB cannot take place. With a fully
localized Higgs, the leading contribution to the soft squared masses
is provided by loops of the electroweak gauge sector, and hence it is
of order $\alpha_W$. This contribution is positive and cannot trigger
EWSB. The usual negative stop/top contributions are however weakened
by the fact that the stop masses are itself a one loop effect
(predominantly generated by gluon/gluino loops) and, hence, are
effectively two loop contributions $\sim \alpha_t\alpha_s$.  As has
been shown in Ref.~\refcite{Barbieri:2002sw}, this contribution is to
weak for EWSB to occur.  There are essentially two ways out of this
dilemma: either the top/stop sector or the Higgs sector is taken to
propagate in the bulk. In the former case the stops feels SUSY
breaking at tree level and the negative contribution to the Higgs
squared mass is enhanced, in the latter case the Higgs soft scalar
mass matrix can possess vanishing or even tachyonic eigenvalues.
We should mention that most models do not yield the MSSM at low
energies. This is due to the fact that the SUSY-breaking and
compactification scales are the same, and unless the SS-parameter is
taken to very small values, there is no regime where the theory can be
formulated as a 4d supersymmetric Standard Model with soft breaking
terms. In other words, there is no mass gap between the superpartners
and the KK-partners.

The scenario with a delocalized Higgs sector is realized, for
instance, in the model of Refs.~\refcite{PQ}
$-$\refcite{DPQ}.Evaluating Eq.~(\ref{masses}) at $\omega=0$, we see
that the supersymmetric Higgs sector consists of a tower of chiral
superfields $H^m_u$ and $H^m_d$ with masses $m=|n\pm \tilde \omega|$
with $n\geq 0$ integer.\footnote{We give the masses in units of $1/R$
and without loss of generailty always assume the twist parameters
$\omega$ and $\tilde \omega$ to lie in the interval $[0,1/2]$.}
SS-SUSY breaking splits the scalar masses of each level as
$|n\pm\tilde \omega\pm\omega|$, with the corresponding eigenstates
$h^m_u\pm {h^m_d}^\dagger$.  A massless mode can be achieved at tree
level by choosing $\omega=\tilde\omega$, which actually constitutes a
flat direction of the tree level potential.  As has been shown in
Ref.~\refcite{DPQ}, radiative corrections can then successfully
trigger EWSB. Had the Higgs been localized, the tree level masses
would be controlled by the $\mu$-parameter and EWSB would be
inpossible to achieve. Due to the fact that the massless mode is a
flat direction at tree level, the Higgs mass turns out to be rather
light ($\lesssim 110$ GeV for the most favourable value of $\omega$).
The most efficient way to raise the Higgs mass is thus to delocalize
the matter sector, or parts of it.

A model of this kind has been proposed in
Ref.~\refcite{Barbieri:2000vh}, albeit with a quite different Higgs
sector that contains only one Higgs doublet.  The reason why this
works is that one can chose the boundary conditions such that an
up-type Higgs survives at $y=0$ and a down-type one at $y=\pi R$:
Choosing $r_0=1\,,\ r_\pi=-1$ in Eq.~(\ref{BCs2}) leaves $\mathcal H$
and $\mathcal H_c$ at $y=0$ and $y=\pi R$ respectively. Assigning
$Y=\frac{1}{2}$ to the whole hypermultiplet, one can write the up type
Yukawa couplings at $y=0$ and the down type ones at $y=\pi R$. In the
matter sector, this requires at least the electroweak doublets to
propagate in the bulk.  The spectrum for the Higgs sector is given by
Eq.~(\ref{bos+-}) with $M'=0$. The supersymmetric spectrum thus
consists of chiral superfields $H^m_u$ and $H^m_d$ with $m$ taking
half-integer values.  In Ref.~\refcite{Barbieri:2000vh} the value
$\omega=\frac{1}{2}$ was chosen, which again leaves a massless Higgs
field at tree level. As has been pointed out,\cite{Ghilencea:2001bw}
although there is no zero mode anomaly in this model, there do appear
localized anomalies at the two boundaries. This anomalies can be
canceled by the introduction of a Chern-Simons term in the
bulk. However, along with these anomalies, quadratically divergent FI
terms are generated at the branes. As mentioned in
Sec.~\ref{interval}, they can be removed by a redefenition of the
field $\Sigma^Y$, which however leads to UV-sensitive bulk mass-terms
and spontaneous quasi-localization.\cite{Ghilencea:2001bw} A careful
analisys of this model, including the effects of arbitrary mass terms
for the matter multiplets in the bulk was performed in
Ref.~\refcite{Barbieri:2002sw,BMP}. Also, the effect of a second Higgs
hypermultiplet, canceling the quadratic divergence, was
incorporated. In the absence of any bulk masses, this Higgs sector is
then equivalent to the model of Ref.~\refcite{DPQ} described in the
previous paragraph with the special value $\omega=\frac{1}{2}$.  It
was found that if the matter sector is completely delocalized, the
tachyonic contributions of the top/stop sector are too strong, thus
resulting in unstable $D$-flat directions. This problem had previously
been realized in Ref.~\refcite{AHNSW}. The authors suggested a
nonrenormalizable quartic term in the superpotential in order to
stabilize the potential at large VEVs. Another possibility is to move
away from the value $\omega=1/2$, as realized, e.g., in
Ref.~\refcite{Delgado:2001si}.\footnote{See discussion of that model
below.}
On the other hand, it was also shown that if the matter sector is
completely localized (for instance, by assigning large bulk masses),
EWSB does not take place, as the top/stop contributions are now
two-loop and cannot overcome the positive one-loop electroweak
ones.\footnote{It should be mentioned that the authors of
Ref.~\refcite{DPQ} still find EWSB to occur at $\omega=1/2$, although
it results in a very small (and phenomenologically unacceptable) Higgs
mass. For such a marginal breaking, the uncertainties in their large
logarithm approximation are probably too significant to definitely decide
whether EWSB takes place, and a precision two-loop calculation as in
Ref.~\refcite{BMP} is needed.  On the other hand, for $\omega<1/2$,
EWSB is not marginal and the simplified treatment of
Ref.~\refcite{DPQ} is justified.} However, slightly delocalizing the
top sufficiently enhances its one loop contribution to the Higgs mass
and EWSB is found to be possible, with the degree of localization
typically in the range $\epsilon\sim 5\%$ and the Higgs mass around
$110 - 125$ GeV for compactification scales $1/R\sim 6-12$ TeV.  The
strong UV sensitivity of the Yukawa couplings in this model requires a
rather low cutoff, $\Lambda\sim (2-3)M_c$,\cite{BMP} above which the
top Yukawa coupling quickly becomes nonperturbative.

The scenario with a delocalized top/stop sector and localized Higgs
doublets is realized, for instance, in the model of
Ref.~\refcite{Delgado:2001si}. In this model, the electroweak singlets
were assumed to be the only bulk matter fields.  Notice that the trace
over the hypercharge of the electroweak singlets vanishes separately,
so no FI term and, hence, no hidden UV sensitivity is generated.  A
$\mu$ parameter has to be introduced in the brane
superpotential. Radiative EWSB is then triggered by the right handed
top/stop sector. Taking $\mu$ and $\omega$ as free parameters of the
model, $\tan \beta$ and the compactification scale are fixed by
the minimization procedure. One typically finds a large value for
$\tan \beta\sim 40$. For such large values of $\tan\beta$ and at
values of $\mu\lesssim 600$ GeV, the SM-like Higgs turns out to be
heavier than the non SM-like one, which has a mass approximately equal
to the pseudoscalar mass, $m_A$. LEP bounds on the latter thus
translate into lower bounds on $\mu$, and in turn on the SM-like Higgs
mass $m_h$. This leads to a rather heavy Higgs, with an
$\omega$-dependent lower bound of approximately $m_h\gtrsim$ 145 GeV.

\section{The MSSM with a Quasi Localized Higgs}
\label{DGQ}

In Sec.~\ref{interval} we have pointed out that the relative UV
insensitivity of the SS-mechanism can be preserved, even with a bulk
Higgs sector, if the latter consists of two hypermultiplet doublets
with brane and bulk masses appropriately chosen such as to produce
boundary conditions that do not generate quadratically or linearly
divergent boundary FI terms. The natural size for the bulk masses lies
somewhere between the fundamental and the compactification scale,
which in turn are typically seperated by a factor $10 - 100$. Once the
bulk mass exceeds the compactification scale, quasi localization of
the lightest modes sets in quickly, while the heavy KK-modes 
decouple from the light spectrum. The minimal setup of this kind would
have the matter sector completely localized. As pointed out before, an
exact localization of the Higgs fields would not result in successful
EWSB. However, for finite values of the localization parameter
$\epsilon$, the lightest Higgs modes do feel SS-SUSY breaking, as
their wave function leeks into the bulk and is sensitive to the SUSY
breaking boundary conditions at the distant brane. Clearly, this
sensitivity can only be of order $\epsilon$, as it must vanish in the
limit of exact localization (i.e.~$\epsilon\to 0$).  In this section
we will outline the model proposed in Ref.~\refcite{Diego:2006py},
which accomplishes EWSB in a very natural manner (i.e.~without large
fine-tuning), and leads to a quite unique superpartner spectrum. Due to
the suppression of the SUSY breaking in the Higgs sector and the
non-interference with the higher KK modes, a description in terms of
the standard MSSM Higgs sector is possible at low energies, although
it has to be borne in mind that above the compactification scale the
theory is modified in an essential way and very different from the
MSSM.

The reason why EWSB can work is that it is possible to generate
tachyonic soft mass terms at tree level that are comparable in size
with both the electroweak one-loop as well as the Yukawa controlled
two-loop contributions. All of them are suppressed w.r.t.~the
compactification scale by loop factors or factors of $\epsilon$.  The
only unsuppressed term is the supersymmetric mass them, i.e.~the $\mu$
parameter. We choose boundary conditions of the form Eq.~(\ref{BCs}),
as well as $M'=0$. Note that this leaves one up and one down-type
chiral Higgs superfield at each brane.  In the limit of unbroken SUSY,
there are two quasilocalized chiral multiplets with mass given by
Eq.~(\ref{locmass})
\be
\mu^2=(1-c_0^2)M^2+ \mathcal O(\epsilon^2)\,,
\ee
where $c_0=\cos (2\pi \alpha_0)$ parametrizes the ``angle'' between
the bulk and brane mass matrices, $c_0=\vec r_0\cdot \vec p$. The bulk
mass scale $M$ is of the order of the compactification scale or
higher, i.e.~in the multi TeV region.  In order to get a $\mu$-term of
roughly the electroweak size and to avoid large cancellations, we
require the angle $\alpha_0$ or equivalently $s_0=\sin(2\pi \alpha_0)$
to be small.

The fact that the $\mu$-term and the soft terms arise at
different orders in the $\epsilon$ expansion can be traced back to the
following fact. Notice that both boundary and bulk mass matrices
preserve $U(1)_H$ subgroups of the global $SU(2)_H$, generated by
$\vec r_f\cdot \vec\sigma$ and $\vec p\cdot\vec\sigma$ respectively.
For $\vec r_0=\pm \vec p$ (corresponding to $s_0=0$) the surviving
$U(1)$ at $y=0$ and the $U(1)$ in the bulk coincide, this symmetry
being broken only by the mismatched $U(1)$ at $y=\pi$. The zero modes
feel this breaking through their wavefunctions, which are, however,
suppressed at $y=\pi$ as $\sim \epsilon$. Hence we expect $\mu\sim
\epsilon^2$ when $s_0=0$ as it can be checked from the $\epsilon$
expansion of fermionic mass eigenvalues. When $s_0\neq0$, the breaking
of the $U(1)$ at $y=0$ is really felt to $\mathcal O(1)$ as it occurs
even for infinitesimally small $y>0$ and hence the $\mu$-term is
unsuppressed.  On the other hand, SUSY is broken \`a la
Scherk-Schwarz, which can be interpreted as a mismatch of the
surviving boundary $U(1)_R$ subgroups of the $N=2$ $SU(2)_R$
automorphism group in the bulk. Again, the zero-mode wave-functions
feel this only to $\mathcal O(\epsilon)$, and the corresponding soft
terms are suppressed as $m^2\sim \epsilon^2$.

The mass eigenvalues for broken SUSY can be inferred from
Eq.~(\ref{masses}).  Since, as just explained, the soft terms are
expected to be $\epsilon$-suppressed with respect to $M_c$ and the
KK-masses, there is a regime where a description in terms of the 4d
MSSM is valid.  Therfore for small $\epsilon$ the MSSM mass Lagrangian
\be 
\mathcal L_{\rm mass}=-(\mu^2+m^2_{H_{u}})\,|H_u|^2
-(\mu^2+m^2_{H_{d}})\,
|H_d|^2+m^2_3\left(H_u\cdot H_d+h.c.\right)
\label{masas}
\ee
can be used. It is thus convenient to translate back the mass
eigenvalues into the soft mass terms of Eq.~(\ref{masas}). With the
additional requirement $s_0\sim \mathcal O(\epsilon)$ one
finds~\cite{Diego:2006py}
\be
m_{H_u}^2=m_{H_d}^2= 4  M^2 \sin^2(\pi\omega)
(1-\tan^2(\pi\tilde \omega))\ \epsilon^2+\dots,
\label{tree}
\ee
\be
m_{3}^2=4  M^2
\sin(2\pi\omega)\tan(\pi\tilde \omega) \
\epsilon^2+\dots,
\ee
while the $\mu$-term is given by
\be
\mu^2 = s_0^2 M^2+\dots
\label{treeb}
\ee
Here, the ellipsis stands for terms suppressed by higher powers of
$\epsilon$ and/or $s_0$.  To these tree level soft masses one has to
add the radiative corrections.  The squark masses will be dominated by
the contribution from the gluinos, which is given by~\cite{ADPQ}
\be
\Delta m_{\tilde t,\tilde b}^2=
\frac{2\,g_3^2}{3\pi^4}\, M_c^2\,
f(\omega)\,,
\label{mstop}
\ee
where the function $f(\omega)$ is defined by
\be 
f(\omega)\equiv \sum_{k=1}^\infty \frac{\sin(\pi k
\omega)^2}{k^3} \,.
\label{fomega}
\ee
Electroweak gauginos provide a radiative correction to the
slepton and Higgs masses as
\be
\Delta^{(1)} m_{H_u}^2=\Delta^{(1)} m_{H_d}^2=
\frac{3g^2+g^{\prime\,2}}{8\pi^4}\,M_c^2
f(\omega)\,.
\label{one}
\ee
Furthermore there is a sizable two-loop contribution to the soft
mass-terms of the Higgs, as well as to the quartic coupling, coming
from top/stop loops with the one-loop generated squark masses given by
Eq.~(\ref{mstop}). This contribution can be estimated in the large
logarithm approximation by just plugging the one-loop squark masses in
the one-loop effective potential generated by the top/stop
sector.\cite{DPQ} For the sake of this paper, where EWSB will not be
marginal (as we will see later) it is enough to consider the effective
potential in the large logarithm approximation, which yields the
two-loop corrections to the Higgs masses
\be
\Delta^{(2)} m_{H_u}^2=\frac{3 y_t^2}{8 \pi^2}\Delta m_{\tilde
t}^2 \, \log \frac{\Delta m_{\tilde t}^2}{\mathcal Q^2}
\,,
\label{twot}
\ee
\be
\Delta^{(2)} m_{H_d}^2=\frac{3 y_b^2}{8 \pi^2}\Delta m_{\tilde
b}^2 \, \log \frac{\Delta m_{\tilde t}^2}{\mathcal Q^2}
\,,
\label{twob}
\ee
where the renormalization scale should be fixed to the scale of SUSY
breaking, i.e.~the gaugino mass $\omega M_c$.\cite{DPQ} Notice that
the corrections from the bottom sector are also considered, which
would only be relevant for large values of $\tan \beta.$
 
Finally, the leading two-loop corrections to the quartic self coupling
of $H_u$ and $H_d$ in the potential 
\be
\Delta V_{\rm quartic}= \Delta\gamma_u |H_u|^4+\Delta\gamma_u |H_u|^4 
\ee
are given by
\be
\Delta\gamma_{u}=\frac{3y_t^4}{16\pi^2}
\log\frac{\Delta m_{\tilde t}^2+m_t^2}{m_t^2}\,,
\ee
\be
\Delta\gamma_{d}=\frac{3y_b^4}{16\pi^2}
\log\frac{\Delta m_{\tilde b}^2+m_b^2}{m_b^2}\,,
\ee
where $m_t$ and $m_b$ are the the top and bottom quark masses
respectively.

Electroweak symmetry breaking can now occur in our model in a very
peculiar and interesting way. The tree-level squared soft masses
$m_{H_u,H_d}^2$ given in Eq.~(\ref{tree}) are suppressed by the factor
$\epsilon^2$ and therefore, for values of $M\sim M_c$ they can be
comparable in size to the one-loop gauge corrections
$\Delta^{(1)}m_{H_u,H_d}^2$ given by Eq.~(\ref{one}).  Furthermore,
the tree-level masses $m_{H_u,H_d}^2$ are negative for values of
$\tilde\omega>1/4$ and then there can be a (total or partial)
cancellation between the tree-level and one-loop contributions to the
Higgs masses. Under extreme conditions they can even cancel,
$m_{H_u,H_d}^2+\Delta^{(1)}m_{H_u,H_d}^2\simeq 0$, in which case the
negative two-loop corrections $\Delta^{(2)}m_{H_u}^2$ will easily
trigger EWSB. On the other hand, in the limit of exact localization of
the Higgs fields $\epsilon\to 0$ the tree-level masses will vanish and
the one-loop gauge and two-loop top/stop corrections have to compete,
which will make the EWSB marginal, as pointed out in
Ref.~\refcite{Barbieri:2002sw,BMP}. These simple arguments prove that
there is a wide region in the space of parameters
$(\omega,\tilde\omega,\epsilon)$ where EWSB easily happens without
much fine-tuning of these parameters.\footnote{A more thorough
treatment of the fine tuning issue can be found in
Ref.~\refcite{Diego:2006py}, where it was shown that the amount of
fine tuning is 10\% (4\%) for $M_c=6.6\ (10)$ TeV.} Of course EWSB
also depends on the Higgsino mass $\mu$ and on the compactification
scale $M_c$ (or equivalently on the gluino mass as it happens in the
MSSM) and we will be concerned about the possible fine-tuning in those
mass parameters.

What other phenomenological requirements could possibly restrict the
parameter space? One constraint arises if we require the
existence of a stable Dark Matter (DM) particle. Note that gaugino
masses are given by $\omega M_c$, while Higgsinos are controlled by
the much smaller $\mu$-parameter. We thus expect the neutralino to be
almost entirely Higgsino-like, with a mass basically given by
$\mu$. Clearly, the charged sleptons must be heavier than the
Higgsinos. Their mass is controlled again by the gaugino mass $\omega
M_c$, although it is smaller by a loop factor. The $\mu $-term also
implicitly increases with $M_c$ through the minimization conditions,
but for smaller $\tilde\omega$ the absolute value of the soft mass
terms in Eq.~(\ref{tree}) decreases, which in turn allows for a
smaller $\mu$.  The requirement that the neutralino be lighter than
the charged sleptons thus favours the region $\omega >\tilde \omega$.
As we will see below, a DM abundance consistent with recent WMAP
results~\cite{WMAP} points towards a rather large compactification
scale $\sim 50$ TeV.

\begin{figure}[htb]
\includegraphics[width=.45\linewidth]{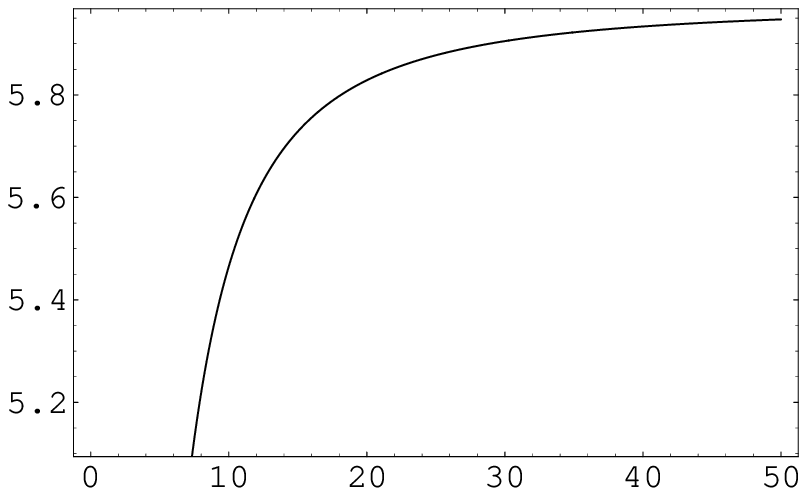}
\hspace{0.5cm}
\includegraphics[width=.47\linewidth]{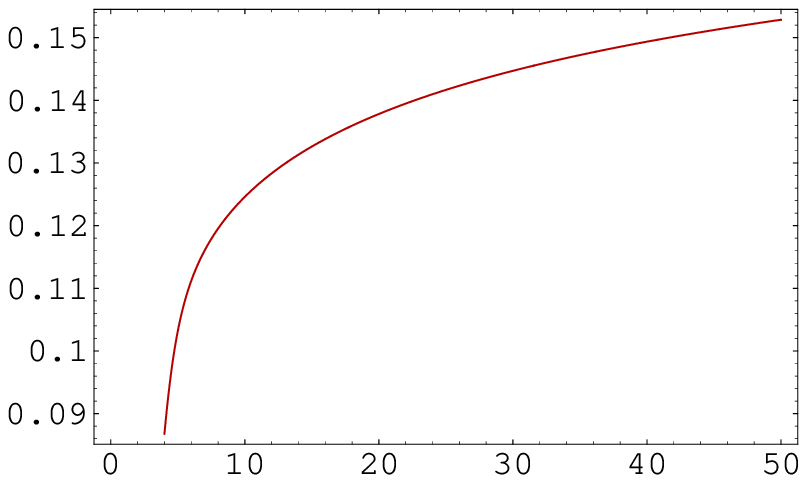}\\\vspace{0.25cm}
\includegraphics[width=.45\linewidth]{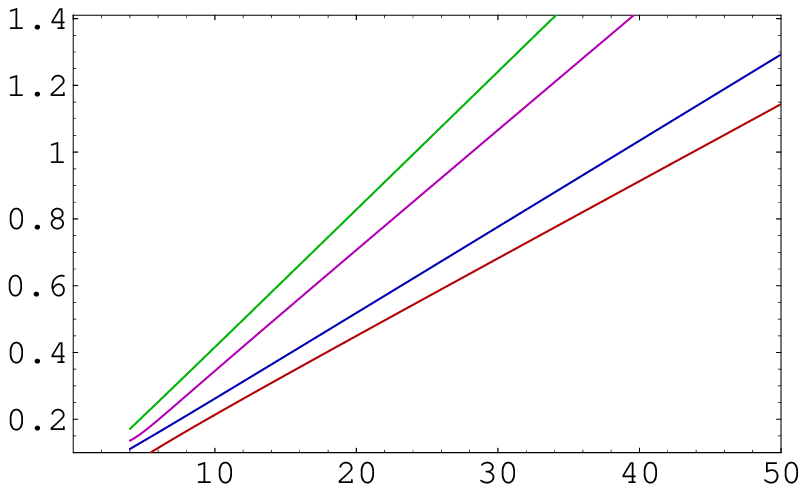}
\hspace{0.8cm}
\includegraphics[width=.45\linewidth]{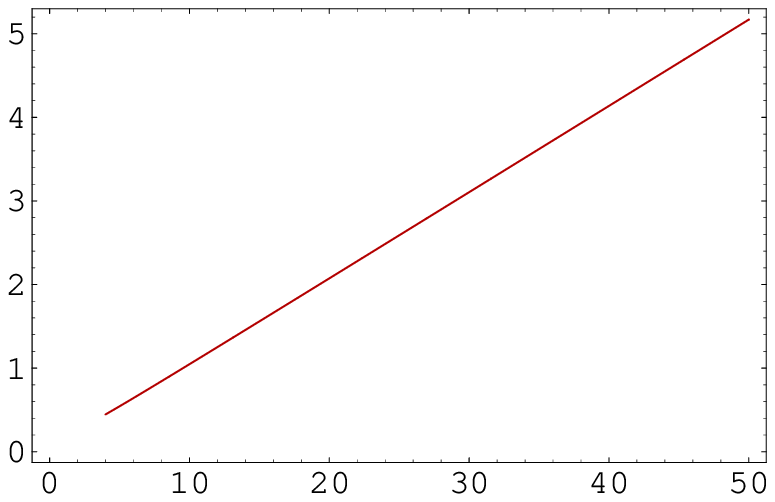}
\caption{\it Predictions for the case $\omega=0.45$,
$\tilde\omega=0.35$, $M=1.65M_c$ as a function of the compactification
scale. Upper left panel: $\tan\beta$. Upper right panel: the SM-like
Higgs mass $m_h$. Lower left panel, from top to bottom the lines
correspond to the masses of: left-handed sleptons $m_{\tilde\ell_L}$,
heavy neutral Higgs (with a mass approximately equal to the
pseudoscalar mass) $m_H\simeq m_A$, right-handed sleptons $m_{\tilde
e_R}$ and neutralinos $m_{\chi^0}\simeq \mu$. Lower right panel: the
squark masses $m_{\tilde q}$. All masses are in TeV.}
\label{espectro}
\end{figure}

We can now solve the minimization conditions for a suitable values of
$(M/Mc,\omega,\tilde\omega)$, which will give us two predictions,
$\tan\beta$ and $\mu$ as functions of the only left free parameter,
$M_c$. We thus can compute the entire mass spectrum as a function of
the compactification scale. The result is displayed in
Fig.~\ref{espectro}.  In the Higgs sector all masses are obtained from
the effective potential where the one-loop corrections to the quartic
couplings are included. The mass of the SM-like Higgs is then computed
with radiative corrections to the quartic couplings considered at the
one-loop level. The SM-like Higgs mass easily satisfies the experimental bound
$m_{h^0}>114.5$ GeV for $M_c>6.5$ TeV. The LSP is the Higgsino-like
with mass $\sim\mu$.  Electroweak precision observables also put lower
bounds on $M_c$ (see e.g.~Ref.~\refcite{Delgado:1999sv}). For the
particularly chosen model the $\chi^2(M_c)$ distribution has a minimum
around $M_c\simeq 10.5$ TeV and one deduces $M_c>4.9$ TeV at 95\% c.l.

The most salient prediction of this model is the ratio of the
squark and slepton masses. As the Higgs sector is effectively
localized, to leading order in perturbation theory these masses are
generated entirely by gauge/gaugino loops. They are finite and
calculable due to the nonlocal nature of SS breaking.  The leading
contribution to the stop and sbottom masses were already given in
Eq.~(\ref{mstop}), adding all gauge contributions one finds
\be 
(m_{\tilde
q_L},\,m_{\tilde u_R},\,m_{\tilde d_R},\,m_{\tilde \ell_L},\,m_{\tilde
e_R}) =(0.110,\,0.103,\,0.102,\,0.042,\,0.025)\sqrt{f(\omega)}M_c \,.
\label{relacion}
\ee
The ratios of the masses are independent on $M_c$ and $\omega$, and
are simply calculated from the couplings and group theoretic
invariants. Recall that similar relations are known from gauge
mediation models (see
Ref.~\refcite{Giudice:1998bp} for a review). There however scalar
masses are generated at the two loop level and hence different ratios
apply. Depending on the size of the messenger scale and other details
of the model, these relations can receive important corrections from
renormalization group (RG) running.  In our case we expect RG effects
to be small, as the high scale ($M_c$) is at most two orders of
magnitude above the low scale (the actual masses). Other small
correcions are expected from higher loops as well as
EWSB.\footnote{There are also one-loop generated $A$ terms.\cite{ADPQ,DPQ}
The effect on the stop mass is approximately $(\Delta^Am_{\tilde
t}/m_{\tilde t})^2\sim v/M_c$.}

\begin{figure}[htb]
\begin{center}
\includegraphics[width=8cm]{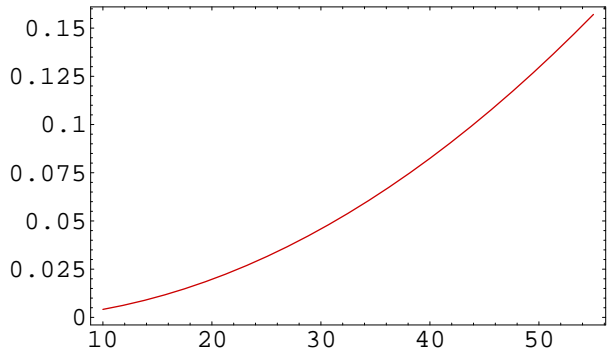}
\caption{\it $\Omega_{\tilde{\chi}^0} h^2$ as a function of $M_c$ (in TeV) for
the model presented in Fig.~\ref{espectro}.}
\label{Omega}
\end{center}
\end{figure}

Finally in the considered class of models where the neutralino is the
LSP and $R$-parity is conserved the lightest neutralino is the
candidate to Cold Dark Matter. The prediction of
$\Omega_{\tilde{\chi}^0} h^2$ can be obtained using the DarkSUSY
package~\cite{Gondolo:2002tz} and can also be approximated by the
expression~\cite{Giudice:2004tc}
\be
\Omega_{\tilde{\chi}^0} h^2\simeq 0.09\left(\mu/{\rm TeV}\right)^2
\ee
In the particular model of Fig.~\ref{espectro} the prediction of
$\Omega_{\tilde{\chi}^0} h^2$ is given in Fig.~\ref{Omega}

Recent WMAP results~\cite{WMAP} imply that
$0.114<\Omega_{\tilde{\chi}^0} h^2<0.134$. As one can see from
Fig.~\ref{Omega} this range in $\Omega_{\tilde{\chi}^0} h^2$ points
towards the range
$49\, {\rm TeV}<M_c<53\,{\rm TeV}$.\footnote{Such large values for $M_c$ require an increased fine-tuning.} Then for a value of $M_c\sim 50$
TeV the density of Dark Matter agrees with the recent results obtained
from WMAP. Notice that for such large values of $M_c$ the neutralinos
are almost Dirac particles. However the non-Dirac character is spoiled
by $\mathcal O(m_W/M_{1/2}) m_W\sim 300$ MeV which is enough to avoid
the strong limits on Dirac fermions that put a lower bound on the
non-Diracity around 100 KeV.\cite{diracity,ABDQT} On the other hand the
WMAP range for $M_c$ implies, in the gravitational sector, gravitino
masses $m_{3/2}\gtrsim 10$ TeV (depending on the value of the SS
parameter $\omega$) are such that gravitinos decay early enough to
avoid cosmological troubles and thus solving the longstanding
cosmological gravitino problem.\cite{Weinberg:1982zq}

\section{Conclusions}

The interplay of Extra Dimensions and supersymmetry can help to
construct realistic models of electroweak symmetry
breaking. Supersymmetry breaking by the SS-mechanism leads to finite
and calculable soft mass terms, controlled by the compactification
scale. In order to generate strong enough supersymmetry breaking in
the Higgs sector, either the top/stop sector or the Higgs sector must
propagate in the bulk of the Extra Dimension, although they may be
quasi-localized to a high degree due to the presence of bulk mass
terms.  We have presented a model that uses a quasilocalized Higgs to
successfully trigger EWSB. Both $\mu$-term and tachyonic soft mass
terms are present at tree level and are seen to have a geometric
origin though the boundary conditions. Moreover, the leading
tree-level, one- and two-loop contributions to the soft squared masses
can naturally be of the same order, leading to a rather modest
fine-tuning in this model.  The characteristics of the model are a
compactification scale in the multi TeV range ($M_c\gtrsim 5$ TeV),
heavy (universal) gauginos $M_{1/2}=\omega M_c$ (with $\omega=.25 -
.5$) and a characteristic and calculable ratio of squark/slepton
masses, Eq.~(\ref{relacion}). The LSP is a neutralino (which is almost
pure Higgsino). It is a good DM candidate for higher values of
$M_c\sim 50$ TeV.

\section*{Acknowledgments}

I would like to thank my collaborators D.~Diego and M~Quir\'os. This
work was supported by grants NSF-PHY-0401513, DE-FG02-03ER41271 and
the Leon Madansky Fellowship, as well as the Johns Hopkins Theoretical
Interdisciplinary Physics and Astrophysics Center .

\end{document}